\documentclass[aps,prd,onecolumn,showpacs,amssymb]{revtex4}
\usepackage{graphicx,bm,color}
\usepackage{amsmath}
\usepackage{amssymb}
\usepackage{amsfonts}
\usepackage[dvips]{epsfig}
\newcommand{\be}{\begin{equation}}
\newcommand{\ee}{\end{equation}}
\newcommand{\bea}{\begin{eqnarray}}
\newcommand{\eea}{\end{eqnarray}}
\newcommand{\beaa}{\begin{eqnarray*}}
\newcommand{\eeaa}{\end{eqnarray*}}

\newcommand{\nn}{\nonumber \\}

\allowdisplaybreaks[4]
\begin{document}

\title{Classical and quantum cosmology of minimal massive bigravity}
\author{F. Darabi}\email{ f.darabi@azaruniv.edu}
\author{ M. Mousavi}\email{ mousavi@azaruniv.edu}

\affiliation{Department of Physics, Azarbaijan Shahid Madani University, Tabriz, 53714-161 Iran}

\begin{abstract}
In a Friedmann-Robertson-Walker (FRW) space-time background we study the classical cosmological models in the context of recently proposed theory
of nonlinear minimal massive bigravity. We show that in the presence of perfect
fluid the classical field equations  acquire contribution from the massive graviton as a cosmological term which is positive or negative depending on the dynamical competition between two scale factors
of bigravity metrics. We obtain the classical field equations for flat and open universes in the ordinary and Schutz representation of perfect
fluid. Focusing on the Schutz representation for flat universe, we find classical
solutions exhibiting singularities at early universe with vacuum equation of state. Then, in the Schutz representation, we  study the quantum cosmology for flat universe and derive the Schrodinger-Wheeler-DeWitt
equation. We find its exact and wave packet solutions and discuss on their properties to show that the initial singularity in the classical solutions
can be avoided by quantum cosmology. Similar to the study of Hartle-Hawking no-boundary proposal in the quantum cosmology of de Rham, Gabadadze and Tolley (dRGT) massive gravity, it turns out that the mass of graviton predicted by quantum cosmology of the minimal massive bigravity is large at early universe. This is in agreement with the fact that at early universe  the
cosmological constant should be large.
\\
\\
Keywords: Massive bigravity, Wheeler-DeWitt equation, Schutz representation

\end{abstract}

\pacs{98.80.-k, 98.80.Qc, 04.50.-h}

\maketitle

\section{Introduction \label{Sec1}}
Since 1916, Einstein's general relativity theory (GR)  \cite{1} has explained the majority of the phenomena related to gravity. Newtonian dynamics is reproduced from GR in the weak field limit that makes observable predictions such as the solar system tests  beside the bending of light by massive objects up
to a very high precision. The scalar curvature of a metric tensor beside the Einstein-Hilbert action in GR theory, proposes a geometrical interpretation of the gravitation. In other words, Einstein's field equations represents the interplay between the space-time geometry and the matter. However, beyond the solar system scales, different astrophysical observations have raised questions that remained unanswered in the framework of GR. These open questions include the cosmological constant problem   \cite{2},  the advent of an invisible dark matter component in the universe \cite{3}, and the recently observed acceleration of the universe
 \cite{4}. Finding explanations  for these unanswered questions convinced
people  to study the possibility of modifying GR theory. From a more theoretical point of view, searching for the alternatives of GR is motivated by string theory, the well-known candidate for a quantum theory of gravity. In the context of field theory, GR is presenting non-linear self interactions of a massless spin-2 particle. From this point of view,
modifying GR can make the spin-2 particle massive. Constructing a consistent theory that describes a massive spin-2 particle is an old challenge. Fierz and Pauli in 1939 presented linearized massive spin-2 field fluctuation \cite{5}. After 30 years, van Dam, Veltman and Zakharov discovered that the helicity-zero mode of the massive spin-2 field does not decouple in the zero-mass limit \cite{6,7}. This is called vDVZ discontinuity and implies that even a very small graviton mass has serious influence on the gravitational interactions between sources.  After a while, Vainshtein presented the Vainshtein mechanism showing that around massive sources the linear approximation (Fierz-Pauli mass term) loses its validity below the Vainshtein radius which leads to the requirement for a nonlinear extension of the Fierz-Pauli mass term \cite{8}. In 1979, Boulware and Deser showed that any nonlinear extension of Fierz-Pauli theory would exhibit a ghost instability \cite{9,10}, and this finding left the massive gravity theory  without any significant progress for about 40 years. Eventually, in 2010, de Rham, Gabadadze and Tolley (dRGT) found that it is possible to construct a theory of ghost-free  nonlinear massive gravity \cite{11,12}. They showed that this theory becomes free of ghost just in a certain decoupling limit which was not enough to guarantee the consistency of the theory. Later, Hassan and Rosen proved the absence of the Boulware-Deser ghost in a Hamiltonian constraint analysis \cite{13,14}. dRGT model is describing a nonlinearly interacting massive spin-2 field in flat space in which two metric tensor components are playing  roles, one as dynamical $g_{\mu\nu}$ and one as non-dynamical $f_{\mu\nu}$ (reference metric). As a result, in the cosmological context of massive gravity theory, it has been shown that the flat Friedmann-Robertson-Walker (FRW) universe dose not exist \cite{15}, but this is not supported by the recent
observational results. However, open FRW solutions are allowed \cite{16} but they are involved with the problems of strong coupling  \cite{17} and ghostlike instabilities \cite{18}. Attempting to obviate the problems with
flat FRW solutions, guided Hassan and Rosen to extend the massive gravity theory, beyond the dRGT setup, to a theory with two dynamical symmetric tensors $g_{\mu\nu}$ and  $f_{\mu\nu}$ as foreground and background metrics, respectively, having a completely symmetric role.  They called this theory as {\it massive bigravity theory} and showed that the corresponding Hamiltonian description
is a ghost-free bimetric theory containing nonlinear interactions of a massless and massive spin-2 field in a dynamical background \cite{19}. It is worth to mention that bimetric theory was first introduced by Isham, Salam and Strathdee \cite{19'} to describe some features of strong interactions and it was later followed and renovated by Damour and Kogan in order to address new physics scenarios \cite{20'}. Apparently, this modified model covers the massive gravity and consider two metrics thoroughly in a symmetric way that annihilate the aether-like concept of reference  metric in massive gravity
\cite{May}. Massive gravity cosmology have been studied in Refs \cite{20,21,22,22',a,b,c,d,e}, meanwhile in bigravity model some regular cosmological solutions have been derived \cite{23,24,24',a1,b1,c1,d1}.

{ At present there is an unsolved question that whether our universe contains closed, flat or open spatial three-geometries. A flat universe
looks like the ordinary three-dimensional space we experience around us. In contrast, the spatial sections of a closed universe looks like three-dimensional  spheres with a very large but finite radius. An open universe looks like an infinite hyperboloid.  Curvature of space is important
if the universe is assumed to be created in an inflationary
state. Because the closed models have finite size
they have generally been thought to be most relevant for
quantum cosmology as they present finite action leading to a non-vanishing
nucleation probability. In fact, Atkatz and Pagels have already shown that
only a closed universe can arise via quantum tunneling \cite{AT}. However, strong evidences from present cosmological observations
favour  a flat or open universe. Therefore, if we believe that quantum mechanics
is the fundamental theory of whole nature, the study of quantum
cosmology for flat or open universe is inevitable and deserves
 more investigation. In 1983, Hawking and Hartle developed a theory of quantum cosmology known as the {\it ``No Boundary Proposal''} \cite{H-H}. We know that the application of path integral to cosmology involves a sum over four dimensional geometries that have boundaries matching onto the initial and final three geometries. The Hartle-Hawking proposal  simply avoids the initial three geometry and only includes four dimensional geometries that match onto the final three geometry. Therefore, path integral is interpreted as giving the probability of a universe with certain properties  being created from nothing. In practice, the calculation of probabilities in quantum cosmology using the path integral is  rather difficult and semiclassical approximation has to be used, where one argues that most of the four dimensional geometries considering in the path integral give rise to very small contributions to the path integral. Indeed, the path integral can be calculated by just considering a few specific geometries known as {\it ``Instantons''} having  considerably large contributions. An instanton describes the spontaneous appearance of a universe from nothing. People have found different types of instantons that can provide the initial conditions for the realistic universes. The first attempt to find an instanton within the context of the `no boundary' proposal was made by Hawking and Moss \cite{H-M}. The Hawking-Moss instanton describes the creation of an eternally inflating universe with {\it closed} spatial three-geometries. Later in 1987, the Coleman-De Luccia instanton was discovered to overcome the limitation of only having closed spatial three-geometries
\cite{C-L}. They showed that the false vacuum decay proceeds via the nucleation of bubbles whose interior is an infinite {\it open} universe in which inflation may occur. Hawking and Turok in 1998  proposed a new class of instantons that give rise to open universes, in a similar way to the instantons of Coleman and De Luccia, without requiring the existence of a false vacuum \cite{H-T}. The Hawking-Turok instantons
essentially make use of the fact that in de Sitter space all curvatures are equivalent. Different slicing
of the 5-dimensional de Sitter hyperboloid correspond to
different curvatures  when considered as a 4-dimensional
model \cite{B-D}. But if we are to require an open or flat universe based
on observations, instead of closed one, the production of such universes using Coleman-De Luccia instantons or Hawking-Turok instantons would seem
a rather convoluted procedure. Indeed, since the non-closed universes can also be compact by topological identifications \cite{J-P} it seems
possible and plausible to study the quantum cosmology of a universe of arbitrary curvature, instead of starting with a closed universe that can later by quantum tunneling create locally open regions. Specifically, it is discussed in \cite{J-H} that although one is generally interested in quantum cosmology of closed universes, but one may retain all three values of the curvature in this study.  
Such point of view has also been followed in detail by Coule and Martin \cite{C-M}.
They have shown that in the flat or open universe, the superpotential of  Wheeler-DeWitt equation is significantly modified, namely the forbidden region (Euclidean region)  goes away and the qualitative behavior of a typical wavefunction differs from that of   closed universe. Because of the absence
of Euclidean nature of the model,   the smooth geometric picture of the Hartle-Hawking ``no boundary proposal'' seems to be lost. Restricting to the {\it Tunneling boundary condition} \cite{T-T}, and applying it  to each of closed, flat
and open universes, it is shown that the quantum cosmology actually favors  the open universe. Considering the above discussions,  it turns out that
unlike closed models which have a forbidden Euclidean region at small scale factors, one can work  with
quantum cosmological models with arbitrary curvature, such as flat and open
universes, whose superpotential of  Wheeler-DeWitt equation does not contain a forbidden Euclidean region. Actually, there are plenty of interesting quantum cosmological models for flat and open universes in which rather than focusing on the closed
universes and calculating the {\it ``creation from nothing''} probability, the Wheeler-DeWitt equation is obtained for flat or open universe and those solutions of Wheeler-DeWitt equation are taken with the criteria of
just having {\it good} asymptotic behaviour in minisuperspace   giving
rise to normalizable states or wave-packets \cite{S-T}. 
 A relevant work to the present paper has also
been recently reported \cite{25} in which the quantum cosmology for the open FRW universe was studied based on the Hamiltonian formalism for massive gravity theory and the corresponding wave packet solutions were obtained. Motivated by this recent study,
in this paper, we shall study the classical cosmology for the flat and open FRW universes as well as quantum cosmology for the flat FRW universe based on the Hamiltonian formalism of the FRW cosmology for massive bigravity theory
\cite{26,27,28,29,30}. In this regard, we use the canonical formulation and quantization of the phase space variables of the minimal massive bigravity model \cite{31}. We derive the Schrodinger-Wheeler-DeWitt
equation and  find its exact and wave packet solutions and then discuss on their properties.}

The organization of this paper is as follows. In section 2, first we review the  minimal bigravity action with two metrics $g_{\mu\nu}$ and $f_{\mu\nu}$ in the presence of perfect fluid and find the point-like Lagrangian where
the graviton's mass plays the role of a cosmological constant. Then, we extract the Hamiltonian density and also the Hamiltonian constraint equations for both metrics $g_{\mu\nu}$ and $f_{\mu\nu}$. In section 3, we obtain the classical solutions, specially in   the Schutz representation of perfect fluid \cite{32,jala}, and  investigate their different aspects such as the effect of graviton's mass as a cosmological constant, in addition to the appearance of the singularities and also the late time accelerated
expansion. In section 4, we  study the quantum cosmology in Schutz representation by extracting the Schrodinger-Wheeler-DeWitt equation for  $g_{\mu\nu}$ and $f_{\mu\nu}$, and find the wave function describing the quantum behavior of the universe. Interpretation of the wavefunction leads to the result that
the graviton's mass at early universe should be large and this may justify
the large cosmological constant at early universe.  The paper ends with a conclusion  in section 5. {Here, we work in units where $c=\hbar=1$.}\\

\section{Point-Like Lagrangian and Hamiltonian constraint in minimal Bigravity Theory \label{Sec2}}

The action of Hassan-Rosen theory named bigravity theory \cite{19} has the following structure
\be
\label{Fbi1}
S_{{\rm bi}}=M_{g}^{2}\int d^{4}x\sqrt{- {\rm det} g}R+M_{f}^{2}\int d^{4}x\sqrt{- {\rm det} f}\tilde{R}+
2m^{2}M_{{\rm eff}}^{2}\int d^{4}x\sqrt{- {\rm det} g}\sum_{n=0}^{4}\beta_{n}e_{n}\left(\sqrt{g^{-1}f}\right)+\int d^{4}x \sqrt{- {\rm det} g}~\mathcal{L}_{m}.
\ee
Here, $g_{\mu\nu}$ and $f_{\mu\nu}$ are two dynamical metric tensors with $R$ and $\tilde{R}$ as the scalar curvatures for tensors $g_{\mu\nu}$ and $f_{\mu\nu}$, respectively, and $\mathcal{L}_{m}~(g,\Phi)$ is the matter
source containing an scalar field $\Phi$. Meanwhile, $m$ is the mass of
graviton and $M_{{\rm eff}}$ is defined as
\begin{align}\label{Fbi2}
\frac{1}{M_{{\rm eff}}^{2}}=\frac{1}{M_{g}^{2}}+\frac{1}{M_{f}^{2}} .
\end{align}
The tensor $\sqrt{g^{-1}f}$ means $\left(\sqrt{g^{-1}f}\right)^{\mu}~_{\rho}\left(\sqrt{g^{-1}f}\right)^{\rho}~_{\nu}=g^{\mu\rho}f_{\rho\nu}=X^{\mu}~_{\nu}$. The trace of this  tensor as $X^{\mu}~_{\mu}$ or $[X]$ helps us to write the following expressions for $e_{n}(X)$'s
\begin{align}\label{Fbi3}
e_{0}(X)=&1,~~e_{1}(X)=[X],~~e_{2}(X)=\frac{1}{2}\left([X]^{2}-[X^{2}]\right),\nn
e_{3}(X)=&\frac{1}{6}\left([X]^{3}-3[X][X^{2}]+2[X^{3}]\right),\nn
e_{4}(X)=&\frac{1}{24}\left([X]^{4}-6[X]^{2}[X^{2}]+3[X^{2}]^{2}+8[X][X^{3}]-6[X^{4}]\right),\nn
e_{k}(X)=&0~~{\rm for}~~k>4.
\end{align}
More simplification motivates us to study the minimal but a non-trivial case as follows \cite{31}

\begin{align}
\label{Fbi4}
S_{{\rm bi}}=&M_{g}^{2}\int d^{4}x\sqrt{- {\rm det} g}R+M_{f}^{2}\int d^{4}x\sqrt{- {\rm det} f}\tilde{R}\nn&
+2m^{2}M_{{\rm eff}}^{2}\int d^{4}x\sqrt{- {\rm det} g}\left(3-{\rm tr}\sqrt{g^{-1}f}+{\rm det }\sqrt{g^{-1}f}\right)+\int d^{4}x \sqrt{- {\rm det} g}~\mathcal{L}_{m},
\end{align}
where use has been made of   (\ref{Fbi3}) as
$$3e_{0}\left(\left(\sqrt{g^{-1}f}\right)^{\mu}~_{\nu}\right)-e_{1}\left(\left(\sqrt{g^{-1}f}\right)^{\mu}~_{\nu}\right)
+e_{4}\left(\left(\sqrt{g^{-1}f}\right)^{\mu}~_{\nu}\right)=3-{\rm tr}\sqrt{g^{-1}f}+{\rm det }\sqrt{g^{-1}f}.$$
Note that considering the non-minimal models leads to quite complicate calculations, whereas in the minimal model the interaction term of two metrics $g_{\mu\nu}$ and $f_{\mu\nu}$ is just obtained by the trace of $\left(\sqrt{g^{-1}f}\right)^{\mu}~_{\nu}$. \\
Let us now obtain the equations of motion   by varying the action (\ref{Fbi4})
with respect to $g_{\mu\nu}$ and $f_{\mu\nu}$, respectively as
\begin{align}
\label{Fbi5}
0=&M_{g}^{2}\left(-R_{\mu\nu}+\frac{1}{2}g_{\mu\nu}R\right)+T_{\mu\nu}\nn&+m^{2}M_{{\rm eff}}^{2}\left\{g_{\mu\nu}\left(3-{\rm tr}\sqrt{g^{-1}f}\right)+\frac{1}{2}f_{\mu\rho}\left(\sqrt{g^{-1}f}\right)^{-1 \rho}~_{\nu}+\frac{1}{2}f_{\nu\rho}\left(\sqrt{g^{-1}f}\right)^{-1 \rho}~_{\mu}\right\},
\end{align}
and
\begin{align}\label{Fbi6}
0=&M_{f}^{2}\left(-\tilde{R}_{\mu\nu}+\frac{1}{2}f_{\mu\nu}\tilde{R}\right)\nn&+m^{2}M_{{\rm eff}}^{2}\sqrt{{\rm det}\left(f^{-1}g\right)}\left\{-\frac{1}{2}f_{\mu\rho}\left(\sqrt{g^{-1}f}\right)^{ \rho}~_{\nu} -\frac{1}{2}f_{\nu\rho}\left(\sqrt{g^{-1}f}\right)^{ \rho}~_{\mu}+f_{\mu\nu}{\rm det}\left(\sqrt{g^{-1}f}\right)\right\}.
\end{align}
As a consequence of the Bianchi identity and also the covariant conservation of $T_{\mu\nu}$, equation (\ref{Fbi5}) leads to the Bianchi constraint for $g_{\mu\nu}$
\be\label{Fbi7}
0=-g_{\mu\nu}\nabla^{\mu}_{g}\left(tr\sqrt{g^{-1}f}\right)+\frac{1}{2}\nabla^{\mu}_{g}\left\{f_{\mu\rho}\left(\sqrt{g^{-1}f}\right)^{-1 \rho}~_{\nu}+f_{\nu\rho}\left(\sqrt{g^{-1}f}\right)^{-1 \rho}~_{\mu}\right\}.
\ee
Similarly, equation (\ref{Fbi6}) gives us the Bianchi constraint for $f_{\mu\nu}$
\be\label{Fbi8}
0=\nabla^{\mu}_{f}\left[\sqrt{{\rm det}(f^{-1}g)}\left\{-\frac{1}{2}\left(\sqrt{g^{-1}f}\right)^{-1\nu}~_{\sigma}g^{\sigma\mu}-
\frac{1}{2}\left(\sqrt{g^{-1}f}\right)^{-1\mu}~_{\sigma}g^{\sigma\nu}+f^{\mu\nu}{\rm det}\left(\sqrt{g^{-1}f}\right)\right\}\right].
\ee
In order to  extract the point-like Lagrangian of this theory, we assume
two dynamical line elements as
\be
\label{Fbi9}
ds_{g}^{2}=-N(t)^{2}dt^{2}+a(t)^{2}\left(\frac{dr^{2}}{1-K r^{2}}+r^{2}d\theta^{2}+r^{2}\sin^{2}\theta d\varphi^{2}\right),
\ee
\be
\label{Fbi10}
ds_{f}^{2}=-M(t)^{2}dt^{2}+b(t)^{2}\left(\frac{dr^{2}}{1-K r^{2}}+r^{2}d\theta^{2}+r^{2}\sin^{2}\theta d\varphi^{2}\right),
\ee
where $N(t)$ and $M(t)$ are the lapse functions, $a(t)$ and $b(t)$ are the scale factors of metrics  $g_{\mu\nu}$ and $f_{\mu\nu}$, respectively, and $K$ is the space curvature {which is assumed to be the same for both metrics \cite{31',32'}}. \\

Upon substitution of these metric coefficients and the definitions of Ricci scalars into the equations (\ref{Fbi3}) and (\ref{Fbi4}), besides the simplification assumption $M_{g}^{2}=M_{f}^{2}=M^{2}_{{\rm eff}}/2$, we can obtain a point-like form for the gravitational Lagrangian in the minisuperspace $\{N,a,M,b\}$ as

\begin{align}
\label{Fbi11}
\mathcal{L}_{{\rm point-like}}=
\frac{a\dot{a}^{2}}{N}-KNa+\frac{b\dot{b}^{2}}{M}-KMb+2 m^{2}\left(N\left(ba^{2}-a^{3}\right)+M\frac{\left(a^{3}-b^{3}\right)}{3}\right).
\end{align}
For the metrics (\ref{Fbi9}) and (\ref{Fbi10}), the Bianchi constraints (\ref{Fbi7}) or (\ref{Fbi8}) equivalently gives

\be\label{Fbi12}
\frac{M}{N}=\frac{\dot{b}}{\dot{a}},
\ee
which is an important result for expediting our next calculations. In order to check the correctness of the relation  (\ref{Fbi11}), we calculate the equations of motion of the variables $M$ and $N$, respectively
\begin{align}
\label{Fbi13}
\frac{\dot{a}^{2}}{N^{2}a^{2}}+\frac{K}{a^{2}}+2m^{2}\left(1-\frac{b}{a}\right)=0,
\end{align}

\begin{align}
\label{Fbi14}
\frac{\dot{b}^{2}}{M^{2}b^{2}}+\frac{K}{b^{2}}+\frac{2m^{2}}{3}\left(1-\frac{a^{3}}{b^{3}}\right)=0.
\end{align}
On the other hand, by inserting the line elements (\ref{Fbi9}) and (\ref{Fbi10}) in the field equations (\ref{Fbi5}) and (\ref{Fbi6}), defining the Hubble parameters $H=\frac{\dot{a}}{Na}$ and $L=\frac{\dot{b}}{b}$, and making use of the Bianchi constraint (\ref{Fbi12}), we can find the following Friedmann equations

\begin{align}
\label{Fbi15}
H^{2}+\frac{K}{a^{2}}+2m^{2}\left(1-\frac{b}{a}\right)=0,
\end{align}

\begin{align}
\label{Fbi16}
3L^{2}+3\frac{KM^{2}}{b^{2}}+2m^{2}M^{2}\left(1-\frac{a^{3}}{b^{3}}\right)=0.
\end{align}
The above equations are exactly the same equations \eqref{Fbi13} and \eqref{Fbi14}
obtained by using the point-like Lagrangian (\ref{Fbi11}),
so this shows the correctness of the point-like Lagrangian.
Comparing \eqref{Fbi15} (and \eqref{Fbi16}) with the standard Friedmann equation
reveals that the squared mass of graviton plays the role of a cosmological
term (provided that $\frac{b}{a}$ takes constant value) which is positive (negative) for $b>a$ and negative (positive) for $b<a$. This is interesting because one can interpret the change from a negative to
a positive cosmological term, due to the dynamical competition between two scale factors,
as a phase transition
from deceleration to acceleration era. In other words, it is possible to
account for the recent acceleration of the universe by this model which accommodates
a desirable sign and value change for the cosmological term
so that the universe can experience an acceleration era after a {\it dynamical}
sign and value change
of the cosmological term from negative to positive.
The dynamical character of the cosmological term, due to the dynamical characters
of two scale factors, may also alleviate the coincidence problem in the sense
that it can make it  possible to have a variable dark energy density with the same order of magnitude of the variable matter density. We will not pursue these interesting issues here and leave them to be reported elsewhere.

Clearly, for the time being we have not considered the matter term in the above Lagrangian.   The momentum conjugate to $a$ and $b$ are obtained as
\be\label{Fbi17}
P_{a}=\frac{\partial \mathcal{L}}{\partial \dot{a}}=\frac{2a\dot{a}}{N},
\ee
and
\be\label{Fbi18}
P_{b}=\frac{\partial \mathcal{L}}{\partial \dot{b}}=\frac{2b\dot{b}}{M},
\ee
respectively. The Hamiltonian can be extracted by Legendre transformation $H=\dot{a}P_{a}+\dot{b}P_{b}-\mathcal{L}$, where the terms  $P_{N}$ and $P_{M}$  do not appear in the Hamiltonian because they are the conjugate momenta of two non-dynamical variables $N$ and $M$ respectively, so we
have \be\label{Fbi19}
H=N\mathcal{H}=N\left(\frac{a\dot{a}^{2}}{N^{2}}+Ka+2m^{2}\left(a^{3}-ba^{2}\right)+\frac{b\dot{b}^{2}}{MN}+\frac{KbM}{N}+2m^{2}\left(\frac{M}{N}\right)
\left(b^{3}-a^{3}\right)\right).
\ee
 Therefore, the Hamiltonian constraint reads as $\mathcal{H}=0$.
At quantum level, the operator form of this constraint is
called Wheeler-DeWitt equation which annihilates the wave function of the universe as $\hat{\mathcal{H}}\Psi=0$.

Now, after the study of geometric sector of the model, we turn to the sector
of matter field. It should be noted that the matter part of the action is thoroughly independent of the modification of the bigravity model by mass term. Thus, we can add the matter part Hamiltonian to the geometric part Hamiltonian (\ref{Fbi19})
 to obtain the total Hamiltonian expression. In order to get the matter part Hamiltonian we use a perfect fluid matter with the equation of state
\be\label{Fbi20}
p=\omega \rho,
\ee
where $p$ is the pressure, and $\rho$ is the  energy density. According to Schutz representation for the perfect fluid \cite{32,33}, the matter part Hamiltonian can be written as
\be\label{Fbi21}
H_{m}=N\frac{P_{T}}{a^{3\omega}},
\ee
where $P_{T}$ is the conjugate momentum corresponding to the dynamical variable
$T$ as the collective thermodynamical parameters of the perfect fluid. In fact, this representation will show its advantages in the study of quantum cosmological part of our model in that a time parameter is constructed naturally
in terms of thermodynamical variables of the perfect fluid. Regarding (\ref{Fbi17}) and (\ref{Fbi18}), we can write
\be\label{Fbi22}
\dot{a}=\frac{NP_{a}}{2a}~~~~{\rm and}~~~~ \dot{b}=\frac{MP_{b}}{2b}.
\ee
 Therefore, the Hamiltonian constraint casts in the following form
 \begin{align}\label{Fbi23}
{\mathcal{H}}=\frac{P_{a}^{2}}{4a^{2}}+K+2m^{2}\left(a^{2}-ba\right)+\frac{P_{T}}{a^{1+3\omega}}+\frac{M}{N}
\left(\frac{P_{b}^{2}}{4ba}
+\frac{Kb}{a}+\frac{2m^{2}}{3}\left(\frac{b^{3}}{a}-a^{2}\right)\right)=0,
\end{align}
where it is seen that the matter is just coupled with the metric $g_{\mu\nu}$ as had been specified in the Lagrangian.  In the next section, we are going to study the classical and quantum cosmologies of this model by considering its Lagrangian and Hamiltonian which have been extracted explicitly in this
section.

\section{Classical Cosmological dynamics \label{Sec3}}

Let us take the classical point of view for the cosmological dynamics of the massive bigravity model including the matter part.  We consider the Friedmann equations including the energy density $\rho$ as

\be\label{Fbi24}
{H^{2}}+\frac{K}{a^{2}}+2m^{2}\left(1-\frac{b}{a}\right)-\frac{\rho}{3M^{2}}=0,
\ee

\be\label{Fbi25}
{H^{2}}+\frac{K}{a^{2}}+\frac{2}{3}m^{2}\left(\frac{b^{2}}{a^{2}}-\frac{a}{b}\right)=0.
\ee
Here, we should remind that although there are several related works about the classical cosmological studies in bigravity that provide an expanded spectrum of cosmological solutions, we are still motivated to seek the behavior of the scale factor in the particular case of massive bigravity theory including
the line elements (\ref{Fbi9}) and (\ref{Fbi10}) in a more transparent way. In this regard, we define the following relations
\be\label{Fbi26}
\frac{b}{a}\equiv \lambda (\tilde{\rho}),~~~~~~~~~\tilde{\rho}\equiv \frac{\rho}{3m^{2}M^{2}},
\ee
which in the minimal massive bigravity leads to some specific cosmological solutions \cite{34}.
Using (\ref{Fbi26}), the Friedmann equations (\ref{Fbi24}) and (\ref{Fbi25}) can be written as
\be\label{Fbi27}
H^{2}+\frac{K}{a^{2}}+2m^{2}\left(1-\lambda\right)=\tilde{\rho}m^{2},
\ee
\be\label{Fbi28}
{H^{2}}+\frac{K}{a^{2}}+\frac{2}{3}m^{2}\left(\lambda^{2}-\lambda^{-1}\right)=0.
\ee
Subtracting two above equations to eliminate $H^{2}$ and $\frac{K}{a^{2}}$ gives the following equation
\be\label{Fbi29}
\tilde{\rho}+2\left(\frac{1}{3}\left(\lambda^{2}-\lambda^{-1}\right)-1+\lambda\right)=0.
\ee
Therefor, we can exactly solve this equation to find, among other two additional complex solutions, the following real expression for $\lambda$

\be\label{Fbi30}
\lambda=-1+
\frac{4-\tilde{\rho}}{2^{\frac{1}{3}}\left(-8+3\tilde{\rho}+
\sqrt{-64+48\tilde{\rho}-15\tilde{\rho}^{2}+2\tilde{\rho}^{3}}\right)^{\frac{1}{3}}}+
\frac{\left(-8+3\tilde{\rho}+
\sqrt{-64+48\tilde{\rho}-15\tilde{\rho}^{2}+2\tilde{\rho}^{3}}\right)^{\frac{1}{3}}}{2^{\frac{2}{3}}}.
\ee
It is fruitful to notice, in the late time behavior of an expanding universe, that $\rho$ generally becomes constant  so-called $\rho_{{\rm vac}}$ which means the vacuum energy density. Obviously, the dimensionless parameter $\lambda$ in (\ref{Fbi30}) approaches a constant value at late times, as well. Thus, we can interpret the Friedmann equations (\ref{Fbi27}) and (\ref{Fbi28}) as the equations which clearly are implying that the late time universe always approaches asymptotically towards a de Sitter or anti-de Sitter universe.
For the early universe limit with $\tilde{\rho}\rightarrow \infty$ we again use the equations  (\ref{Fbi30}) which shows the result $\lambda\rightarrow 0$. Having considered the Friedmann equation for $g_{\mu\nu}$ (\ref{Fbi27}), we
are ended up with an evolution for $H^{2}$ dominated by the very large value of $\tilde{\rho}$ and a rather large cosmological constant, similar to the general relativity. As we mentioned, besides the solution (\ref{Fbi30}) we have two more complex solutions which diverge and so are omitted from the explicit physical class of solutions in the limit $\tilde{\rho}\rightarrow \infty$. Thus, we can deduce that the solution (\ref{Fbi30}) which vanishes in this limit, is the physical one because we may have ordinary general relativity in the very early expansion history of the universe.

In summary, the Friedmann equations in this model support the universe subject
to the ordinary general relativistic Friedmann equations with a cosmological constant proportional to the squared
mass of
gravitons $m^2$, and depending on the parameter of the theory they can describe asymptotically de Sitter or anti-de Sitter universe.

Now, we study the behavior of the physical scale factor $a(t)$ or the corresponding Hubble parameter $H$ to show the details of the above discussion. In doing so, we may insert the $\lambda$ expression  (\ref{Fbi30}) in (\ref{Fbi27}) to extract the behavior of the scale factor with respect to  time. Defining $\lambda^{*}=2\left(1-\lambda\right)-\tilde{\rho}$ as an specified expression which only includes $\tilde{\rho}$, we can write
\be\label{Fbi31}
\frac{\dot{a}^{2}}{N^{2}m^{2}a^{2}}+\frac{K}{a^{2}m^{2}}+\lambda^{*}(\tilde{\rho})=0.
\ee
In order to find physical non-oscillating scale factor, we assume the gauge $N=1$ at late times with constant $\lambda^{*}<0$.
Under this assumption,  we have the following cases for two values of $K$:
\\
\\
\textbf{CASE I}:~~~~$K=0$ \\
\\
We get the following form of modified Friedmann equation in bigravity model
\be\label{Fbi32}
 \dot{a}^{2}+m^{2}a^{2}\lambda^{*}=0,
\ee
which has two following branch of solutions
\be\label{Fbi33}
a(t)=e^{\pm {\rm im}(t-t_{0})\sqrt{\lambda^{*}}}~~~{\rm or}~~~a(t)=e^{\mp {\rm m}(t-t_{0})\sqrt{|\lambda^{*}|}}.
\ee
For a limited time interval at late time, they are two branches of solutions describing de Sitter and anti de Sitter-like solutions. Solutions for a flat space with two opposite behaviors of the scale factor, without any singularity in this limited time interval,  is shown in Figure 1. \\

\textbf{CASE II}:~~~~$K=-1$
\be\label{Fbi34}
a(t)=\pm \frac{\sin\left[m\sqrt{\lambda^{*}}\left(t\pm t_{0}\right)\right]}{m\sqrt{\lambda^{*}}}~~~{\rm or}~~~
a(t)=\pm \frac{\sinh\left[m\sqrt{-\lambda^{*}}\left(t\pm t_{0}\right)\right]}{m\sqrt{-\lambda^{*}}}.
\ee

\begin{figure*}[ht]
  \centering
  \includegraphics[width=2.5in]{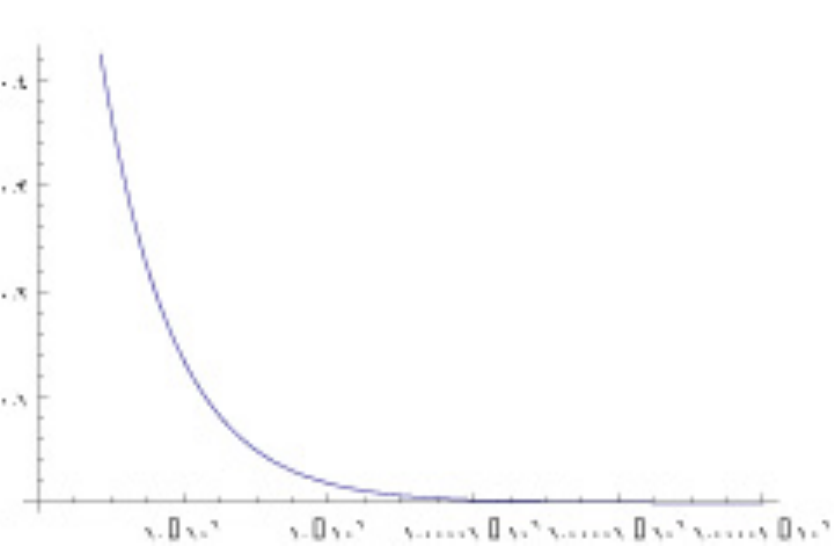}
  \includegraphics[width=2.5in]{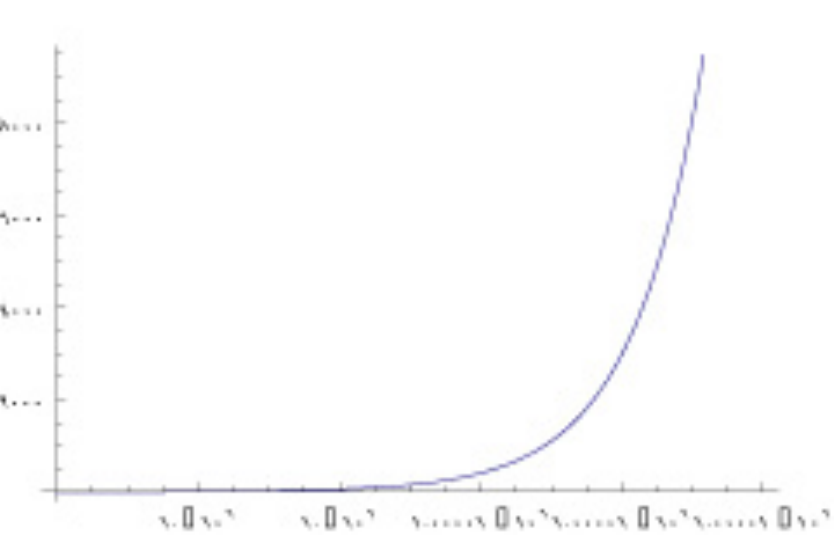}
  \caption{The figures show the evolutionary behavior of the late time universes based on (\ref{Fbi38}) (left and right figures correspond to $e^{-(t-t_{0})}$ and  $e^{+(t-t_{0})}$, respectively). We have used the typical numerical values $t_{0}=10^{6}$
  and $m\sqrt{|\lambda^{*}|}=1$. }

\end{figure*}
It is worth mentioning that the conformal time $\tau$ is equivalent to the cosmic time $t$ since we have chosen the lapse function $N=1$. Anyway, these solutions for an open universe are treated in a manner that with $t_{0}\geq0$ , $t>0$ and  $a(t)\geq0$, they are divided into two branches in which $a_{+}(t)$ and $a_{-}(t)$ are valid for $t\geq-t_{0}$ and $t\leq t_{0}$ respectively, such that $a_{\pm}(t_{0})=0$. In general, the $a_{+}(t)$ evolution begins with a big-bang-like singularity at $t=-t_{0}$ and then goes on its evolution exponentially; at late time of cosmic evolution that is exactly our case, the mass term $m^2$ shows itself as a cosmological constant. The evolution is opposite for the case, $a_{-}(t)$, in a way that the universe gradually decreases its scale factor  from a large initial size at  $t=+\infty$ towards
 a zero size at $t=t_{0}$. Nevertheless, the solutions for $K=-1$ generally differ from the previous ones for $K=0$, because in the present solutions the  initial singularities appear, whereas the de Sitter or (Ads) solution does not carry any initial singularity; in other words, the present solutions  are  similar to the previous ones just at late time. It should be noted that the closed universe case $K=+1$ dose not have any preferable physical solution.

We can follow another approach such that  the matter part is considered as perfect fluid, similar to the previous section. For simplicity in the calculations
of classical and quantum cosmology (see bellow), we consider the Schutz representation \cite{32,33} for the perfect fluid which is coupled with the gravity. In this case, the Hamiltonian (\ref{Fbi21})
describes the dynamics of the system. Thus, we have the following equations of motion for $T$ and $P_{T}$\\
\be\label{Fbi35}
\dot{T}=\left\{T,H\right\}=\frac{N}{a^{3\omega}},~~~~~\dot{P_{T}}=\left\{P_{T},H\right\}=0.
\ee
Considering the above equations, we conclude that choosing $N=a^{3\omega}$ will help us to write

\be\label{Fbi36}
T=t,
\ee
showing the time role of the variable $T$. Before going through the Friedmann equations, we set up the following assumption
\be\label{Fbi37}
\frac{b}{a}=\gamma,
\ee
where $\gamma$ is a function to be obtained later. This assumption gives the opportunity to have an analytical calculation and  getting some explicit results for the scale factor $a(t)$. Taking~ $P_{T}=P_{0}={\rm const},$~ with some similar calculation of the equations (\ref{Fbi27}) and (\ref{Fbi28}) we can write
\be\label{Fbi38}
\frac{\dot{a}^{2}}{a^{6\omega-1}}+\frac{K}{a^{2}}+\frac{P_{0}}{a^{3\left(1+\omega\right)}}+2m^{2}\left(1+\gamma\right)=0,
\ee
\be\label{Fbi39}
\frac{\dot{a}^{2}}{a^{6\omega-1}}+\frac{K}{a^{2}}+\frac{2}{3}m^{2}\left(\gamma^{2}-\gamma^{-1}\right)=0.
\ee
 Considering Eqs.(\ref{Fbi38}), (\ref{Fbi39}), we deduce that $\gamma$ should
 be at most a function of the scale factor $a$ which according to \eqref{Fbi12} means that the lapse coefficient $M$ should be treated as a function of the scale factor $a$, too. In fact, we are interested in this behavior because it can describe a new set of cosmological solutions differing from the de-Sitter or Ads universes. Subtracting equations (\ref{Fbi38}) and (\ref{Fbi39}) gives
\be\label{Fbi40}
-2\left(1+\gamma\right)+\frac{2}{3}\left(\gamma^{2}-\gamma^{-1}\right)=\frac{P_{0}}{m^{2}a^{3\left(1+\omega\right)}}.
\ee
For $K=0$ we can classify the results in two cases:\\
\\
$\bullet$~$\omega=-\frac{1}{3}$:~cosmic string\\
\\
By using (\ref{Fbi40}),  we obtain
\begin{align}\label{Fbi41}
&\gamma=1+\frac{4m^{2}a^{2}+P_{0}}{2^{\frac{1}{3}}\left(12m^{6}a^{6}+3m^{4}a^{4}P_{0}+\sqrt{m^{6}a^{6}
\left(m^{2}a^{2}-P_{0}\right)\left(4m^{2}a^{2}+P_{0}\right)^{2}}\right)^{\frac{1}{3}}}+\nn&
\frac{2^{\frac{1}{3}}\left(12m^{6}a^{6}+3m^{4}a^{4}P_{0}+\sqrt{m^{6}a^{6}
\left(m^{2}a^{2}-P_{0}\right)\left(4m^{2}a^{2}+P_{0}\right)^{2}}\right)^{\frac{1}{3}}}{2^{\frac{2}{3}}m^{2}a^{2}}.
\end{align}
Inserting the above result in the equation (\ref{Fbi38}) helps us to extract the time evolution of the scale factor by means of solving the following reduced equation
\begin{align}\label{Fbi42}
0=&a^{3}\dot{a}^{2}+\frac{P_{0}}{a^{2}}+4m^{2}+\frac{4m^{4}a^{2}+P_{0}m^{2}}{2^{\frac{-2}{3}}\left(12m^{6}a^{6}+3m^{4}a^{4}P_{0}+\sqrt{m^{6}a^{6}
\left(m^{2}a^{2}-P_{0}\right)\left(4m^{2}a^{2}+P_{0}\right)^{2}}\right)^{\frac{1}{3}}}+\nn&
\frac{2^{\frac{2}{3}}\left(12m^{6}a^{6}+3m^{4}a^{4}P_{0}+\sqrt{m^{6}a^{6}
\left(m^{2}a^{2}-P_{0}\right)\left(4m^{2}a^{2}+P_{0}\right)^{2}}\right)^{\frac{1}{3}}}{a^{2}}.
\end{align}
 The behavior of  scale factor with respect to the cosmic time is plotted in Figure 2, as the solution for this equation. Clearly, this is describing a universe in which we have a big-bang singularity at $t=0$ and a positive-valued scale factor increasing monotonically.

\begin{figure*}[ht]
  \centering
   \includegraphics[width=2.5in]{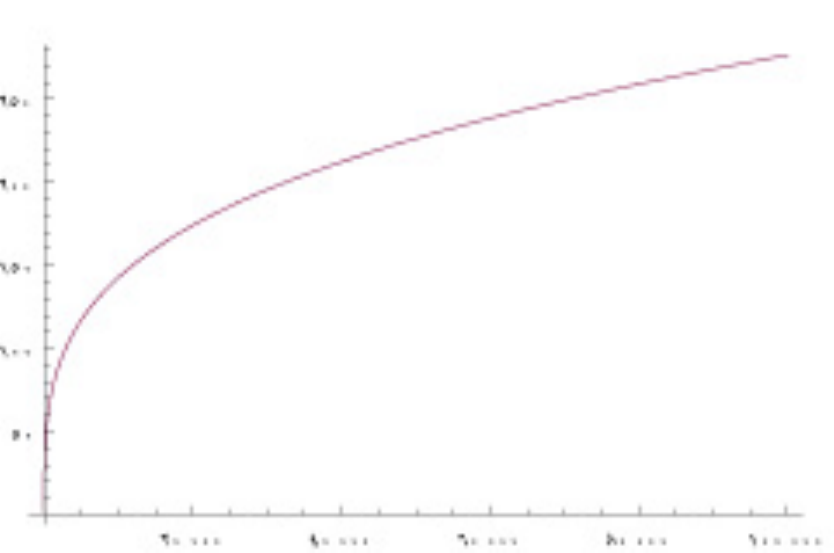}
  \caption{The time evolutionary behavior of the scale factor $a(t)$ according to (\ref{Fbi42}). We have considered the numerical values $m=1$ and $P_{0}=-10^{6}$.  }

\end{figure*}

$\bullet$~$\omega=-1$:~cosmological constant\\ \\
By using (\ref{Fbi40}), we obtain a constant value  $\gamma_{0}$ and so
\be\label{Fbi43}
a^{7}\dot{a}^{2}+P_{0}+2m^{2}\left(1+\gamma_{0}\right)=0,
\ee
which has the following solution
\be\label{Fbi44}
a_{\pm}(t)=\left(\frac{9}{2}\right)^{\frac{2}{9}}\left(\pm\sqrt{-\left(2\left(1+\gamma_{0}\right)m^{2}+P_{0}\right)}
\left(t-t_{0}\right)\right)^{\frac{2}{9}}.
\ee
By means of the relation between the conformal time and the cosmic time $d\tau=a(t)^{-1}dt$ we make the above result more comprehensive as
\begin{eqnarray}\label{Fbi45}
a(\tau)=\left\{
\begin{array}{lll}
\left(\frac{49}{4}\left|\left(2\left(1+\gamma_{0}\right)m^{2}+P_{0}\right)\right|\left(\tau-\tau_{0}\right)^{2}\right)^{\frac{1}{3}}
~~~~~~~~~~~2\left(1+\gamma_{0}\right)m^{2}+P_{0}<0,\\
\\
-\left(\frac{49}{4}\left(2\left(1+\gamma_{0}\right)m^{2}+P_{0}\right)\left(\tau-\tau_{0}\right)^{2}\right)^{\frac{1}{3}}
~~~~~~~~~2\left(1+\gamma_{0}\right)m^{2}+P_{0}>0.
\end{array}
\right.
\end{eqnarray}
Considering the condition $a(\tau)>0$, we  definitely can ignore the case $2\left(1+\gamma_{0}\right)m^{2}+P_{0}>0$ and also we should emphasize that the condition $2\left(1+\gamma_{0}\right)m^{2}+P_{0}<0$ imposes some specific limitation on the parameters $m$ and $P_{0}$ which we disregard here further discussion about its details. Plotting the evolution of the scale factor for this case in Figure 3, we can see a big-bang-like singularity at $\tau=\tau_{0}$  which starts from the singular point and grows gradually with time.

\begin{figure*}[ht]
  \centering
  \includegraphics[width=2.5in]{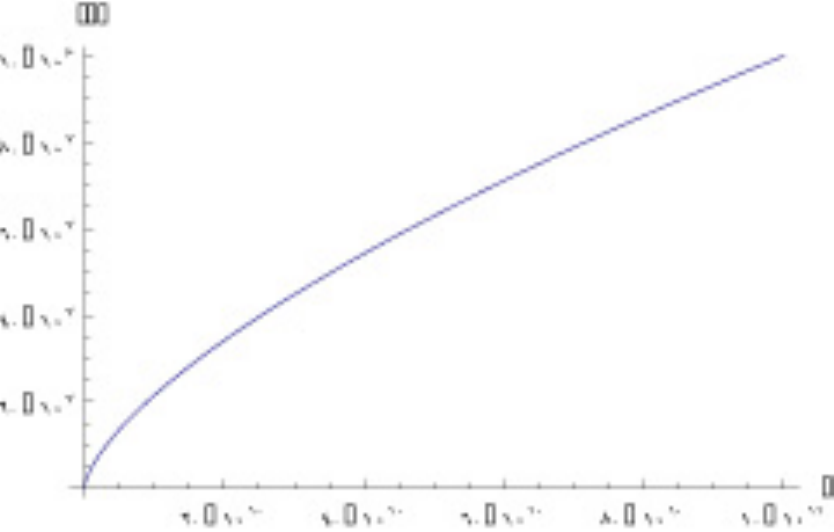}
  \caption{The conformal time evolution of the scale factor $a(\tau)$ obeying (\ref{Fbi45}) with the numerical value consideration $\left|\left(2\left(1+\gamma_{0}\right)m^{2}+P_{0}\right)\right|=100$.  }

\end{figure*}

\section{Quantum Cosmological dynamics \label{Sec4}}
Let us study the quantization of the present model by means of the canonical quantization. To this end, we obtain the Wheeler-DeWitt equation $\hat{\mathcal{H}}\Psi=0$, as the operator version of the Hamiltonian constraint ${\mathcal{H}}=0$, and solve it to find the universe wave function $\Psi$.
By means of (\ref{Fbi23}), the Wheeler-DeWitt equation reads as

\be\label{Fbi46}
\hat{\mathcal{H}}\Psi \left(a,b,T\right)=\left[\frac{\hat{P}_{a}^{2}}{4a^{2}}+K+2m^{2}\left(a^{2}-ba\right)+\frac{\hat{P}_{T}}{a^{1+3\omega}}+
\left(\frac{M}{N}\right)\left(\frac{\hat{P}_{b}^{2}}{4ba}
+\frac{Kb}{a}+\frac{2m^{2}}{3}\left(\frac{b^{3}}{a}-a^{2}\right)\right)\right]\Psi \left(a,b,T\right)=0.
\ee
 We sperate the variables in $\Psi \left(a,b,T\right)$ as follows
 \be\label{Fbi47}
\Psi \left(a,b,T\right)=e^{i E T}\psi(a,b),
\ee
where E is a constant quantity. This gives
rise to\be\label{Fbi48}
\hat{\mathcal{H}}\Psi \left(a,b,T\right)=\left[\frac{\hat{P}_{a}^{2}}{4a^{2}}+K+2m^{2}\left(a^{2}-ba\right)+\frac{E}{a^{1+3\omega}}+
\left(\frac{M}{N}\right)\left(\frac{\hat{P}_{b}^{2}}{4ba}
+\frac{Kb}{a}+\frac{2m^{2}}{3}\left(\frac{b^{3}}{a}-a^{2}\right)\right)\right]\psi \left(a,b\right)=0.
\ee
Let us assume, for simplicity, that  $({b}/{a})=\sigma={\rm cte.}$ Then, the  Bianchi constraint (\ref{Fbi12}) takes the following form
\be\label{Fbi49}
\frac{b}{a}=\sigma=\frac{\dot{b}}{\dot{a}}=\frac{M}{N}.
\ee
Using \eqref{Fbi17}, \eqref{Fbi18} and \eqref{Fbi49}, we find that $P_b=\sigma P_b$ which leads to the following relation between two momentum operators
\be\label{Fbi50}
\hat{P_{b}}={\sigma}\hat{P_{a}}.
\ee
As a result, we have the final form of the  Wheeler-DeWitt equation as
\be\label{Fbi51}
\left[\left(1+\sigma^{2}\right)\frac{\hat{P}_{a}^{2}}{4a^{2}}+\left(1+\sigma^{2}\right)K+\frac{E}{a^{1+3\omega}}
+2m^{2}a^{2}\left(1-\frac{4}{3}\sigma+\frac{\sigma^{4}}{3}\right)\right] \psi \left(a\right)=0.
\ee
This is a direct consequence of the coupling between the matter source and the gravitation $g_{\mu\nu}$ which states that the dynamics of two metrics $g_{\mu\nu}$ and $f_{\mu\nu}$ are identified  and so the two-variable system is mostly simplified in favor of a one-variable system including just one scale factor $a$. In other words, we are allowed to take $\Psi \left(a,b\right)\equiv \Psi \left(a,\sigma\right)\equiv \psi \left(a\right)$.

Note that this Schutz representation is related directly to the well-known {\it time problem} in quantum cosmology context \cite{35}. Clearly, in (\ref{Fbi46}) the conjugate momentum associated with $T$ appears linearly in the Hamiltonian of the model. By means of  the canonical quantization, we obtain a Schrodinger-Wheeler-De
Witt (SWD) equation in which the matter variable $T$ plays the role of time. It is noticeable that this approach dose not solve completely the time problem in massive bigravity quantum cosmology in a primordial way. However, applying  a perfect fluid instead of a field is somehow leads to the emergence of a Schr¨odinger-type equation with a useful time parameter $T$.
The particular solution of this Schr¨odinger-type equation is going to be fined in the rest. Here we should consider the time ordering rule of the operators $\hat{a}$ and $\hat{P}_{a}$ by means of the relation ${\hat{P}^{2}_{a}}=-a^{-p}\frac{\partial}{\partial_{a}}\left(a^{p}\frac{\partial}{\partial_{a}}\right)$, thus (\ref{Fbi51}) becomes

\be\label{Fbi52}
\left[-\frac{1}{4a^{2}}\left(1+\sigma^{2}\right)\left(\frac{d^{2}}{da^{2}}+\frac{p}{a} \frac{d}{da}\right)+\left(1+\sigma^{2}\right)K+\frac{E}{a^{1+3\omega}}
+2m^{2}a^{2}\left(1-\frac{4}{3}\sigma+\frac{\sigma^{4}}{3}\right)\right]\psi \left(a\right)=0.
\ee
The generic solutions of the above equation for the case $\omega=-1$ and $K=0$ ({Taking $K=0$ is due to the fact that it is in agreement with the current observations}) is as follows
\begin{align}\label{Fbi53}
\psi_{E\sigma}(a)=&
(-1)^{\frac{-p}{6}} 3^{\frac{p-3}{12}} a^{\frac{1-p}{2}}\left(1+\sigma^{2}\right)^{\frac{p-1}{12}}\left(3E+2m^{2}\left(3-4\sigma+\sigma^{4}\right)\right)^{\frac{1-p}{12}}
\left[(-1)^{\frac{1}{6}}3^{\frac{p}{6}}{\rm{I}}\left[\frac{1-p}{6},\frac{2a^{3}
\sqrt{3E+2m^{2}\left(3-4\sigma+\sigma^{4}\right)}
}{3\sqrt{3}\sqrt{1+\sigma^{2}}}\right]\times \right.\nn
& \left. c_{2} {\rm{\Gamma}}\left[\frac{7-p}{6}\right]+(-3)^{\frac{p}{6}}{\rm{I}}\left[\frac{-1+p}{6},
\frac{2a^{3}
\sqrt{3E+2m^{2}\left(3-4\sigma+\sigma^{4}\right)}
}{3\sqrt{3}\sqrt{1+\sigma^{2}}}\right]c_{1} {\rm{\Gamma}}\left[\frac{5+p}{6}\right]\right].
\end{align}
By the choice of the factor ordering value $p=-2$ (the factor $p$ denotes the uncertainty in the choice of operator ordering ) \cite{36,37} we have
\be\label{Fbi54}
\psi_{E\sigma}(a)=c_{1}\cosh \left[\frac{2a^{3}
\sqrt{3E+2m^{2}\left(3-4\sigma+\sigma^{4}\right)}
}{3\sqrt{3}\sqrt{1+\sigma^{2}}}\right]+{\rm ic_{2}sinh}\left[\frac{2a^{3}
\sqrt{3E+2m^{2}\left(3-4\sigma+\sigma^{4}\right)}
}{3\sqrt{3}\sqrt{1+\sigma^{2}}}\right].
\ee
Imposing the boundary condition on these solutions as $\psi\left(a=0\right)=0$, we are led to the following result with $c_{1}=0$.
\be\label{Fbi55}
\psi_{E\sigma}(a)=i c_{2}\sinh\left[\frac{2a^{3}
\sqrt{3E+2m^{2}\left(3-4\sigma+\sigma^{4}\right)}
}{3\sqrt{3}\sqrt{1+\sigma^{2}}}\right].
\ee
It is obvious that the wave function $\psi_{E\sigma}(a)$ does not satisfy the squared integrability condition unless we transform it into a ``{\em Sin}'' function. In doing so, we may impose the constraint $3E+2m^{2}\left(3-4\sigma+\sigma^{4}\right)<0$ so that the wave function becomes oscillatory
\be\label{Fbi56}
\psi_{E\sigma}(a)=- c_{2}\sin\left[\frac{2a^{3}
\sqrt{|3E+2m^{2}\left(3-4\sigma+\sigma^{4}\right)|}
}{3\sqrt{3}\sqrt{1+\sigma^{2}}}\right].
\ee
The eigenfunctions of the Wheeler-DeWitt equation may be written as
\be\label{Fbi57}
\Psi_{E\sigma}\left(a,T\right)=e^{i E T}\psi_{E\sigma}(a).
\ee
In Figure 4, we have plotted the square of the eigenfunction $\psi_{E\sigma}(a)$ \eqref{Fbi55} for typical values of the parameters. It seems from this figure that we have a well-defined wave function in the vicinity of $a=0$ describing a universe emerging out of nothing without any tunneling. This is because there is no potential barrier in the classical Hamiltonian \eqref{Fbi52}
for tunneling effect. It is seen that the squared wave function has  peaks,
in the vicinity of specific and countless values of growing scale factors, which are rapidly decaying as the scale factor grows. If we consider this as an instant
spatial configuration of probability for observing a universe, we realize that the universe is observed with most probability, corresponding to the dominant contribution, at the smallest (nonvanishing) scale factor among other possible
scale factors. This means that although there is no tunnelling from nothing scenario for this universe, however, similar to the tunnelling scenario the universe is born from nothing with most probability, close to $a=0$. In other words,  the quantum cosmology
can mimic the tunneling scenario and avoid the classical initial singularity by supporting the probability for creation of universe from nothing at a desired small nonvanishing scale factor.   By extremizing $|\psi_{E\sigma}(a)|^2$, using (\ref{Fbi56}), it is easily found that the size of initial scale factor corresponding to the dominant contribution
of $|\psi_{E\sigma}(a)|^2$ is determined by the mass of gravitons,  and that a small size newborn universe requires
a large mass  for gravitons. Such large mass for gravitons at early universe
has already been proposed
in the study of Hartle-Hawking no-boundary proposal in dRGT massive gravity
\cite{38}.
They have
 shown that the
contribution from the massive gravity sector can substantially enhance the probability of a large number of e-
folding  for a sufficiently large graviton mass comparable to the Hubble parameter
during inflation, namely $m \gtrsim\ 10^{12}GeV$, and  illustrated possible models to trigger such a large graviton mass
at early universe while it is negligibly small in the present universe.

 We may reach to the most precise description of the wave function, provided that we construct a wave packet for this wave function. In order to obtain the general solution as a wave packet we may have superposition of the eigenfunctions $\Psi_{E\sigma}\left(a,T\right)$ with a suitable Gaussian weight function $e^{-\gamma E^{2}}$ as

\begin{figure*}[ht]
  \centering
   \includegraphics[width=2.5in]{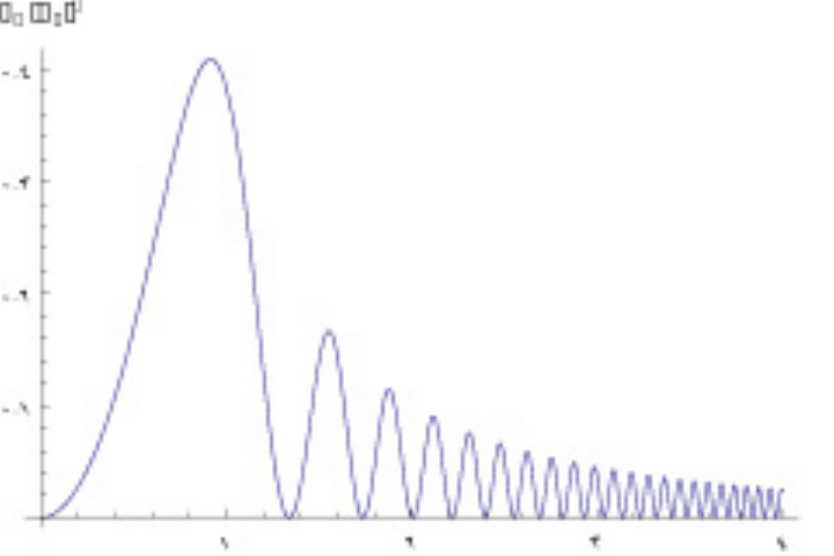}
  \caption{The square of absolute value of the eigenfunction $\psi_{E\sigma}(a)$ for typical values of the parameters.}

\end{figure*}

\be\label{Fbi58}
\Psi_{\sigma_{0}}\left(a,T\right)= \int_{E=0}^{\infty}e^{iET-\gamma E^{2}}  \psi_{E\sigma_{0}}\left(a\right) dE.
\ee

The Gaussian weight function $e^{-\gamma E^{2}}$ and other examples (quasi-Gaussian and shifted Gaussian) are extensively used in quantum mechanics as a tool to obtain localized states. These types of weight functions are concentrated about a specific value of their argument and fall off swiftly away from that central point. Multiplying the Gaussian weight function by the obtained wave function $e^{i ET}\psi_{E\sigma}(a)$, we finally find from (\ref{Fbi58}) a Gaussian-like behavior for the wave packet which is localized about some special values of its argument $a$.
Since the above integral is too complicate  to extract an analytical closed form for the wave function, we have just plotted the squared  wave function $|\Psi_{\sigma_0}\left(a,T\right)|^{2}$ in Figure 5, where $\sigma_{0}=4.7$.
The figure indicates that the wave packet spreads out along both the
spatial axes
$a$ and temporal axes $T$.
 \begin{figure*}[ht]
  \centering
   \includegraphics[width=3in]{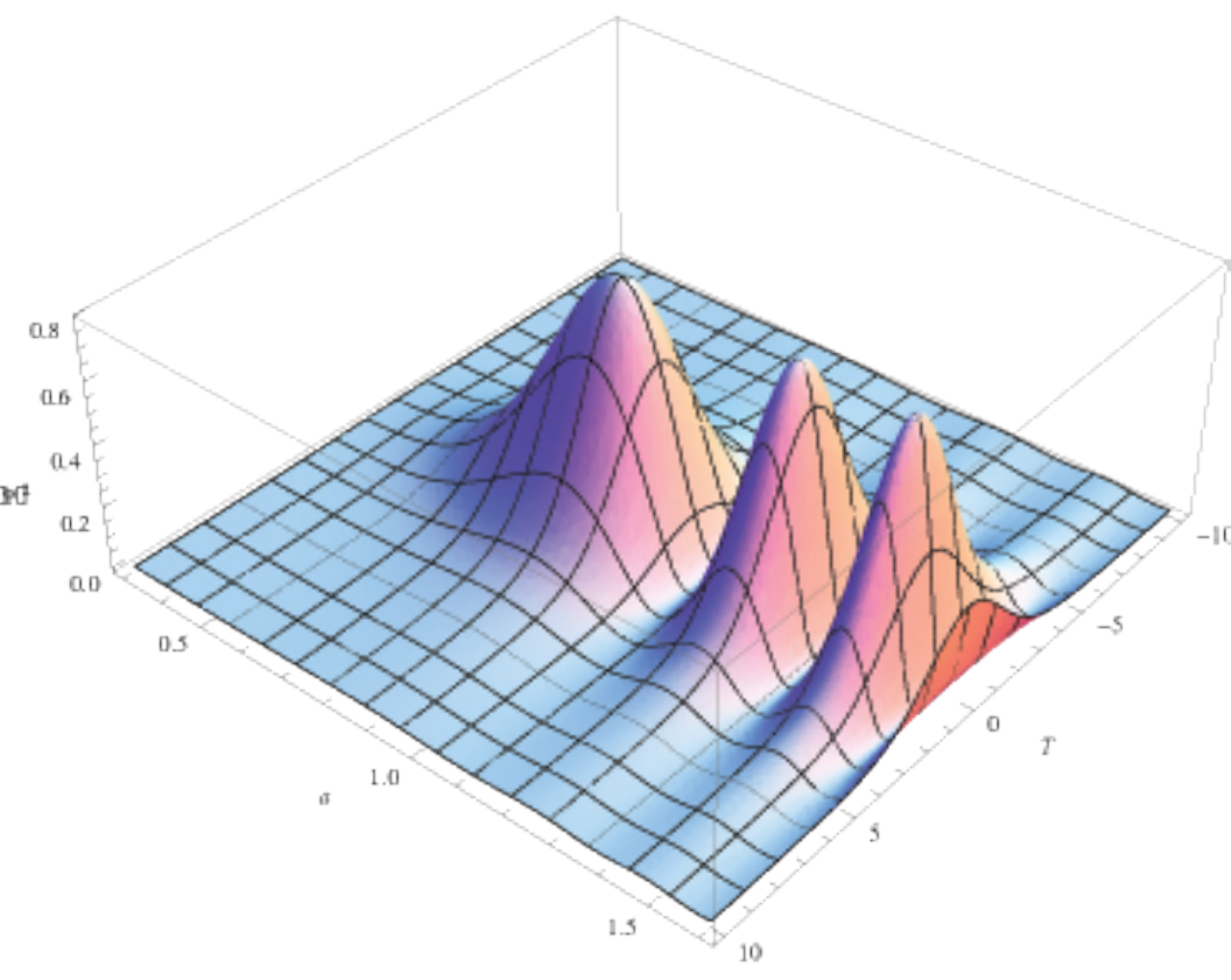}
  \caption{The squared wave function $|\Psi_{\sigma=4.7}\left(a,T\right)|^{2}$ with the numerical assumptions $m^{2}=\gamma=1$.}
\end{figure*}

\section{Conclusions\label{Sec6}}
In this paper, we have applied the recently proposed nonlinear massive bigravity theory to a FRW cosmological model. Using the Hamiltonian formalism for bigravity theory we have studied the classical and quantum cosmological behaviors of the particular massive bigravity model so called minimal bigravity. It is
shown that the classical field equations  receive contribution from the massive gravitons as a cosmological term whose value can change dynamically from negative to positive value, depending on the competition between two scale factors
of bigravity metrics. Such a dynamical cosmological term is capable of accounting
for the recent acceleration of the universe as well as resolving
the coincidence problem, to which we shall address elsewhere. We have presented the classical cosmological solutions for the cases where the universe contains a perfect fluid  with the equation of state parameters
$\omega=-1$ (cosmological constant), and $\omega=-\frac{1}{3}$ (cosmic string). For the general energy density $\rho$, we have found some solutions of two Friedmann equations in flat ($K=0$) and open ($K=-1$) cases. In the flat case, we have obtained two branches of solutions in which there are no singularities. Similarly, in the open universe we have obtained two branches of solutions which are analyzed in details leading to the similar results as those of the flat case except that in this case we have initial singularities.
 For the energy density with Schutz representation of perfect fluid, applied for $\omega=-\frac{1}{3}$ (cosmic string), we have found a universe in which there is a singularity, and then we plotted its scale factor evolution in Figure 2. We have also studied another case $\omega=-1$ (cosmological constant) and found that the situation is similar to the previous case except for the scale factor which evolves more rapidly than that of the cosmic string, and then we plotted its evolution in Figure 3.

 We have investigated the quantization of the minimal bigravity model via the method of canonical quantization in a two variable minisuperspace including two scale factors. Due to the presence of interaction term between two scale factors in the massive bigravity action and using the Bianchi constraint, we could reduce the two variable minisuperspace into one variable minisuperspace including just one scale factor, which is preferred to be $``a(t)"$. Having used Schutz representation for the perfect fluid, beside a particular gauge choice $N={a^{-3}}$ we have introduced a parameter playing the role of time in a Schrodinger-Wheeler-DeWitt equation. In the presence of matter for a particular vacuum case $\omega=-1$ and $K=0$, corresponding to early universe, we have solved the Schrodinger-Wheeler-DeWitt equation exactly and constructed the wave packet corresponding to this solution, numerically. We have found that although there is no tunnelling from nothing scenario for this {\it one variable quantum cosmology}, however,  this quantum cosmology
can avoid the classical initial singularity by supporting the probability for creation of universe from nothing at a  nonvanishing  scale factor. Moreover,
it is shown that the demand for a small initial scale factor for the newborn universe, requires a large mass for gravitons.
Such a large mass of graviton (comparable to the Hubble parameter
during inflation, namely $m \sim 10^{12}$ GeV) at early universe is in agreement
with the fact that the cosmological constant should be large at early universe. Actually, the same result has already been proposed
in the study of Hartle-Hawking no-boundary proposal for the quantum
cosmology of de Rham, Gabadadze and Tolley (dRGT) massive gravity \cite{38}
where the authors have
given two reasons for why the graviton can have large value at early universe while it is negligible today. We may add another reason based the application
of uncertainty
principle on the universe as a single quantum system. The very small size of universe at early times confines the
range of gravitons
to be very small, at most, of the size of universe. According to the uncertainty
principle $\Delta P \Delta X\sim \hbar$, such mediating particles with a small range $\Delta X$ should have a large momentum $\Delta P$,
and this may justify the large mass for gravitons at early universe. As the universe expands, the range of gravitons becomes larger and, according to the uncertainty
principle, the mass of such mediating gravitons becomes smaller.
 In other words, if we assume the universe with the age $\Delta t \sim
H^{-1}$, the mass of gravitons, according to the uncertainty
principle $\Delta E \Delta t\sim \hbar$, should be of the size $m\sim\hbar H/c^2$ which is in agreement with the present prediction on the graviton's
mass and also the prediction of \cite{38} where it is shown that the mass
of graviton at early universe was comparable (in the natural units $\hbar=c=1$) to the Hubble parameter during inflation.
The identification of graviton's mass with the cosmological term and the reduction of graviton's mass by time evolution of the universe may be considered as a solution of the cosmological constant problem.
\section*{Acknowledgments}

We would like to thank B. Vakili, K. Atazadeh and S. Jalalzadeh for many useful discussions
and constructive
comments.
\appendix


\begin{thebibliography}{9}
\bibitem{1} A. Einstein, Annalen, Phys. 49 (1916) 769 [Annalen Phys. 14 (2005) 517].
\bibitem{2} S. Weinberg, Rev. Mod. Phys. 61 (1989) 1.
\bibitem{3} F. Zwicky, Helv. Phys. Acta. 6 (1933) 110.
\bibitem{4}S. Perlmutter et al., Astrophys. J. 517 (1999) 565.
\bibitem{5} M. Fierz and W. Pauli, Proc. Roy. Soc. Land. A 173 (1939) 211.
\bibitem{6} H. Van Dam and M. J. G. Veltman, Nucl. Phys. B22 (1970) 397.
\bibitem{7} V. I. Zakharov, JETP Lett. 12 (1970) 312 [Pisma zh. Eksp. Teor. Fiz 12 (1970) 447].
\bibitem{8} A. I. Vainshtein, Phys. Lett. B 39 (1972) 393.
\bibitem{9} D. G. Boulware and S. Deser, Phys. Lett. B 40 (1972) 227.
\bibitem{10} D. G. Boulware and S. Deser, Phys. Rev. D 6 (1972) 3368.
\bibitem{11} C. de Rham and G. Gabadadze, Phys. Rev. D 82 (2010) 044020 [arXiv:1007.0443].
\bibitem{12} C. de Rham, G. Gabadadze and A. J. Tolley, Phys. Rev. Lett. 106 (2011) 231101 [arXiv:1011.1232].
\bibitem{13} S. F. Hassan and R. A. Rosen, Phys. Rev. Lett. 108 (2012) 041101 [arXiv:1106.3344].
\bibitem{14} S. F. Hassan and R. A. Rosen, JHEP 1204 (2012) 123 [arXiv:1111.2070].
\bibitem{15} G. D'Amico, C. de Rham, S. Dubovsky, G. Gabadadze, D. Pirtskhalava and A. J. Tolley, Phys. Rev. D 84 (2011) 124046 [arXiv:1108.5231].
\bibitem{16} A. E. Gumrukcuoglu, C. Lin and S. Mukohyama, JCAP 1111 (2011) 030 [arXiv:1109.3515].
\bibitem{17} N. Khosravi, G. Niz, K. Koyama and G. Tasinat. JCAP 1308 (2013) 044 [arXiv: 1305. 4950].
\bibitem{18} A. De Felice, A. E. Gumrukcuoglu and S. Mukohyama, Phys. Rev. Lett. 109 (2012) 171101  [arXiv:1206.2080].
\bibitem{19} S. F. Hassan and R. A. Rosen, JHEP 1202 (2012) 126 [arXiv:1109.3515].
\bibitem{19'} C. J. Isham, A. Salam and J. A. Strathdee, Phys. Rev. D 3 (1971) 867.
\bibitem{20'} T. Damour and I. I. Kogan, Phys. Rev. D 66 (2002) 104024 [arXiv:hep-th/0206042].
\bibitem{May}A. Schmidt-May, M. von Strauss, Journal of Physics A: Mathematical and Theoretical, Topical Review, 49, DOI:10.1088/1751-8113/49/18/183001 [arXiv:1512.00021].
\bibitem{20} J. Kluson [arXiv:1209.3612];\\
             C. de Rham, G. Gabadadze, L. Heisenberg and D. Pirtskhalava, Phys. Rev. D 83 (2011) 103516  [arXiv:1010.1780];\\
             S. F. Hassan, A. Schmidt-May and M. Von Strauss, Phys. Lett. B 715 (2012) 335  [arXiv: 1203.5283];\\
             E. N. Saridakis, Class. Quant. Grav. 30 (2013) 075003  [arXiv:1207.1800].
\bibitem{21} J. Kluson, Phys. Rev. D 86 (2012) 044024  [arXiv:1204.2957].
\bibitem{22} S. F. Hassan and R. A. Rosen, JHEP 1204 (2012) 123  [arXiv:1111.2070].
\bibitem{22'}E. N. Saridakis, [arXiv:1207.1800].
\bibitem{a}Yi-Fu. Cai, C. Gao and E. N. Saridakis, [arXiv:1207.3786].
\bibitem{b}Yi-Fu Cai, D. A. Easson, C. Gao and E. N. Saridakis, Phys. Rev. D 87 (2013) 064001.
\bibitem{c}G. Leon, J. Saavedra and E. N. Saridakis, [arXiv:1301.7419].
\bibitem{d}Yi-Fu Cai, F. Duplessis and E. N. Saridakis, Phys. Rev. D 90 (2014) 064051.
\bibitem{e}Yi-Fu Cai and E. N. Saridakis, Phys. Rev. D 90 (2014) 063528.
\bibitem{23} D. Comeli, M. Crisostomi, F. Nesti, L. Pilo, JHEP 1203 (2012) 067 , Erratum-ibid. 1206 (2012) 020  [arXiv:1111.1983].
\bibitem{24} M. Von Strauss, A. Schmidt-May, J. Enander, E. Mortsell and S. F. Hassan, JCAP 1203 (2012) 042 [arXiv:1111.1655].
\bibitem{24'}S. Nojiri and S. D. Odintsov, [arXiv:1207.5106].
\bibitem{a1}S. Nojiri, S. D. Odintsov and N. Shirai, [arXiv:1212.2079].
\bibitem{b1}K. Bamba, A. N. Makarenko, A. N. Myagky, S. Nojiri and S. D. Odintsov, JCAP 01 (2014) 008.
\bibitem{c1}K. Bamba, S. Nojiri and S. D. Odintsov, Phys. Lett. B 731 (2014) 257.
\bibitem{d1}S. Nojiri, S. D. Odintsov and V. K. Oikonomou, [ arXiv:1511.06776].
\bibitem{AT}D. Atkatz and H. Pagels, Phys. Rev. D 25 (1982)  2065.
\bibitem{H-H}J. Hartle and S. W. Hawking, Phys. Rev. D 28 (1983) 2960. 
\bibitem{H-M}S. W. Hawking and I. G. Moss, Phys. Lett. B 110 (1982) 35.
\bibitem{C-L}S. Coleman and F. De Luccia, Phys. Rev. D 21 (1980) 3305.
\bibitem{H-T}S. W. Hawking and N. Turok. Phys. Lett. B 425 (1998) 25. 
\bibitem{B-D}N. D. Birrell and P. C. W. Davies, {\it Quantum Fields in
curved space}, Cambridge University press, Cambridge (1982).
\bibitem{J-P}J. P. Luminet, Phys. Report 254 (1995) 135; D. N. Page, in Proceedings of the Banff Summer Institute on Gravitation,
August 1990, eds. R. B. Mann and P. S. Wesson, (World Scientific, Singapore,
1991).
\bibitem{J-H}J. J. Halliwell, in {\it Quantum Cosmology and Baby Universes}, eds. S. Coleman, J. B. Hartle T. Piran and S. Weinberg, (World Scientific, Singapore, 1991), p. 159.
\bibitem{C-M}D. H. Coule and J\'er\^ome Martin, Phys. Rev. D 61 (2000) 063501.
\bibitem{T-T}A. Vilenlin, Phys. Rev. D 27 (1983) 2848.
\bibitem{S-T}C. Kiefer, Nucl. Phys. B 341  (1990) 273, T. Dereli, M. \"Onder and R. W. Tucker, Class. Quantum Grav. 10 (1993) 1425; F. Darabi and H. R. Sepangi, Class. Quantum Grav. 16 (1999) 1565; S. S. Gousheh, H. R. Sepangi, P. Pedram, M. Mirzaei, Class. Quantum Grav. 24 (2007) 4377; B. Vakili, S. Jalalzadeh, H. R. Sepangi, JCAP 0505 (2005) 006; S. Jalalzadeh, F. Ahmadi, H. R. Sepangi, JHEP 0308 (2003) 012; F. Darabi, A. Rastkar, Gen. Rel. \textsf{\textsf{a}}Grav. 38 (2006) 1355; P. Pedram, S. Jalalzadeh, Phys. Rev. D 77 (2008) 123529.
\bibitem{25} B. Vakili and N. Khosravi, Phys. Rev. D 85 (2012) 083529 [arXiv:1204.1456].
\bibitem{26} J. Kluson, (2014) [arXiv:1307.1974].
\bibitem{27} V. O. Soloviev (2014) [arXiv:1410.0048].
\bibitem{28} V. O. Soloviev and Margarita V. Tchichikina (2013) [arXiv:1211.6530].
\bibitem{29} V. O. Soloviev (2015) [arXiv:1505.0084].
\bibitem{30} V. O. Soloviev (2013) [arXiv:1312.5516].
\bibitem{31} S. F. Hassan and R. A. Rosen, JHEP 1107 (2011) 009  [arXiv:1103.6055].
\bibitem{31'} M. Von Strauss, A. Schmidt-May, J. Enander, E. Mortsell and S. Hassan, JCAP 1203 (2012) 042 [arXiv:1111.1655].
\bibitem{32'} Y. Akrami, T. S. Koivisto and M. Sandstad, JHEP 1303 (2013) 099 [arXiv:1209.0457].
\bibitem{32} B. F. Schutz, Phys. Rev. D2 (1970) 2762;\\
             B. F. Schutz, Phys. Rev. D4 (1971) 3559;\\
             V. G. Lapchinskii and V. A. Rubakov, Theor. Math. Phys. 33 (1977) 1076.

\bibitem{jala}P. Pedram, S. Jalalzadeh and S. S. Gousheh, Phys. Lett. B655 (2007) 91;\\
              P. Pedram, S. Jalalzadeh and S. S. Gousheh, Class. Quant. Grav. 24 (2007) 5515;\\
              P. Pedram and S. Jalalzadeh, Phys. Rev. D77 (2008) 123529.
\bibitem{33} A. B. Batista, J. C. Fabris, S. V. B. Goncalves and J. Tossa, Phys. Lett. A 283 (2001) 62 [arXiv:0011102];\\
             F. G. Alvarenga, J. C. Fabris, N. A. Lemos and G. A. Monerat, Gen. Rel. Grav. 34 (2002) 651 [arXiv:0106051];\\
             A. B. Batista, J. C. Fabris, S. V. B. Goncalves and J. Tossa, Phys. Rev. D 65 (2002) 063519 [arXiv:0108053];\\
             B. Vakili, Phys. Lett. B 688 (2010) 129 [arXiv:1004.0306];\\
             B. Vakili, Class. Quantum Grav. 27 (2010) 025008 [arXiv:0908.0998].
\bibitem{34} M. Van Strauss, A. Schmidt-May, J. Enander, E. Mortsell and S. F. Hassan (2011) [arXiv:1111.1655].
\bibitem{35} B. S. DeWitt, Phys. Rev. 160 (1967) 1113.

\bibitem{36} D. He, D. Gao and Q-yu Cai, Phys. Lett. B 784 (2015) 361  [arXiv:1507.06727].
\bibitem{37} D. He, D. Gao and Q-yu Cai, Phys. Rev. D 89 (2014) 083510  [arXiv:1404.1207].
\bibitem{38} M. Sasaki, D-han Yeom, Y-li Zhang, Class. Quant. Grav. 30 (2013)
232001 [arXiv:1307.5948].

\end{thebibliography}
\end{document}